\documentclass[12pt]{article}
\usepackage[centertags]{amsmath}
\usepackage{amssymb}
\usepackage{graphicx}
\usepackage{cite}
\begin{document}

\setcounter{page}{0}
\thispagestyle{empty}

\begin{flushright}
{\small BARI-TH 489/2004}
\end{flushright}

\vspace*{2.5cm}

\begin{center}
{\large \bf Large logarithmic rescaling of the scalar condensate:}  \\[0.4cm]
{\large \bf new lattice evidences}
\end{center}

\renewcommand{\thefootnote}{\fnsymbol{footnote}}

\begin{center}
{%\large
P. Cea$^{1,2,}$\protect\footnote{Electronic address: {\tt
Paolo.Cea@ba.infn.it}},
M. Consoli$^{3,}$\protect\footnote{Electronic address: {\tt
Maurizio.Consoli@ct.infn.it}},
and L. Cosmai$^{1,}$\protect\footnote{Electronic address: {\tt
Leonardo.Cosmai@ba.infn.it}} \\[0.5cm]
$^1${\em INFN - Sezione di Bari, I-70126 Bari,
Italy}\\[0.3cm]
$^2${\em Dipartimento
Interateneo di Fisica, Universit\`a di Bari, I-70126 Bari,
Italy}\\[0.3cm]
$^3${\em INFN - Sezione di Catania, I-95123 Catania,
Italy}
}
\end{center}

\vspace*{0.5cm}

\begin{center}
{%\large
July, 2004}
\end{center}

\vspace*{1.0cm}

\renewcommand{\abstractname}{\normalsize Abstract}
\begin{abstract}
Using two different methods,
we have determined the rescaling of the scalar condensate
$Z\equiv Z_\varphi$ near the critical line of a 4D Ising model.
Our lattice data, in agreement with previous numerical indications,
support
the behavior $Z_\varphi\sim \ln ({\Lambda})$, $\Lambda$ being the ultraviolet cutoff.
This result is
predicted in an alternative description of symmetry breaking where
there are no upper bounds on the Higgs boson mass from `triviality'.
\vspace{1pc}
\end{abstract}

%\begin{keyword}
%lattice\sep spontaneous symmetry breaking \sep Higgs
%\PACS 14.80.Bn, 11.10.-z, 11.15.Ha

%\end{keyword}
%\end{frontmatter}

\section{Introduction}
There are many computational and analytical evidences pointing towards the
`triviality' of $\Phi^4$ theories in $3+1$
dimensions~\cite{Sokal_book,Lang:1993sy} (see also ~\cite{Kenna:2004cm} and
references therein),
though a rigorous proof is still lacking.
Nevertheless these theories continue to be useful
and play an important role for unified model of electroweak interactions.

The conventional interpretation of these theories
extends to any number N of scalar field components and,
when used in the Standard Model, leads to predict a
proportionality relation, $m^2_H \sim g_R v^2_R$,
between the squared Higgs boson mass $m_H^2$
and the square of the
known weak scale $v_R$ (246 GeV) through the renormalized
scalar self-coupling
$g_R \sim 1/{\ln \Lambda}$, $\Lambda$ being the ultraviolet cutoff.
In this picture, the ratio $m_H/v_R$ is a cutoff-dependent quantity
that becomes smaller and smaller when $\Lambda$ is made larger and larger.

By accepting the validity of this
picture, there are important phenomenological implications.
For instance, a precise measurement of $m_H$, say $m_H=760 \pm 21$ GeV,
would constrain the possible values of $\Lambda$ to be smaller
than about 2 TeV thus suggesting the occurrence of `new physics' at that
energy scale.

In an alternative approach~
\cite{Consoli:1994jr,Consoli:1997ra,Consoli:1999ni},
%\cite{Consoli:1994jr,Agodi:1995qv,Consoli:1997ra,Consoli:1999ni},
however, this conclusion is not true.
The crucial point is that
the `Higgs condensate' and its quantum fluctuations undergo different
rescalings when changing the ultraviolet cutoff. Therefore,
the relation between $m_H$ and the physical $v_R$
is not the same as in perturbation theory.

In order to remind the basic issue let us preliminarily observe that in a
broken-symmetry phase the conventional field rescaling cannot be viewed
as an `operatorial statement' between bare and renormalized fields
of the type, say
\begin{equation}
\label{operator}
                     "~\Phi_B(x)= \sqrt{Z} \Phi_R(x)~"
\end{equation}
In fact \cite{Agodi:1995qv} this relation is a consistent short-hand notation
in a theory allowing an
asymptotic Fock representation, as in QED. However, in the presence of
spontaneous symmetry breaking it has no rigorous
basis since the Fock representation exists only for the {\em shifted}
fluctuating field. For this reason, it is the residue of the shifted-field
propagator, say $Z\equiv Z_{\text{prop}}$, that through the
K\'allen-Lehmann representation is related to the normalization of the
single-particle states. In principle,
this quantity is quite unrelated to $Z\equiv Z_\varphi$
the rescaling of the vacuum field (the scalar `condensate') which is
defined through the physical
mass and the zero-momentum susceptibility and
is by no means constrained to be below unity.

To be definite, let us consider a one-component scalar theory and introduce
the bare expectation value
\begin{equation}
\label{vB}
v_B=\langle\Phi_{\text{ latt}}\rangle
\end{equation}
 associated with
the `lattice' field as defined at a locality scale fixed by the ultraviolet
cutoff. Connecting to the stability analysis,
such expectation value represents one of the absolute minima,
say $\varphi_B=\pm v_B$, of the effective potential
$V_{\text{eff}}(\varphi_B)$ of the theory.

Now, by $Z\equiv Z_\varphi$ we denote the rescaling that is needed to obtain
the physical vacuum field
\begin{equation}
\label{vR}
v_R= \frac{v_B}{\sqrt{Z_\varphi}}  \,.
\end{equation}
By {\em physical}, we mean that the second derivative
of the effective potential
$V''_{\text{eff}}(\varphi_R)$ parameterized in terms of the physical field and
evaluated at $\varphi_R=\pm v_R$, is precisely given by
$m^2_H$. Since the second derivative of the effective potential
$V''_{\text{eff}}(v_B)$ is the bare
zero-four-momentum two-point function (the inverse zero-momentum
susceptibility),
this standard definition is equivalent to define $Z_\varphi$ as
\begin{equation}
\label{z1phi}
      Z_\varphi= \frac{m^2_H}{V''_{\text{eff}}(v_B)}=
  m^2_H \chi_2 (0)
\end{equation}
where $\chi_2(0)=1/V''_{\text{eff}}(v_B)$
is the bare zero-momentum susceptibility. Notice that there
is nothing in the above derivation that dictates $Z_\varphi \leq 1$.

On the other hand, assuming
the K\'allen-Lehmann representation for the shifted fluctuating field, that
has a vanishing expectation value and
admits a particle interpretation, one predicts
$0 < Z_{\text{prop}}\leq 1$,
with `triviality' implying $Z_{\text{prop}} \to 1$
when approaching the continuum theory.
Therefore, although in the standard approach one assumes
$Z_\varphi =Z_{\text{prop}} = 1+ {\mathcal{O}} (g_R)$,
up to small perturbative corrections, one should remind the basically
different operative definitions of the two Z's.

For this reason, a different interpretation
of triviality~was proposed (see
Refs.\cite{Consoli:1994jr,Consoli:1997ra,Consoli:1999ni}
and some elder works quoted therein) starting
from the observations that `triviality' does not require the effective
potential to be
a trivially quadratic function of $\varphi$. Thus, the theory can be
`trivial', in a technical sense, but non-trivial in a physical sense
and the two things can coexist provided all interaction effects can be
re-absorbed, in the continuum limit,
into the vacuum structure and the physical mass of a massive free field.

This requirement leads to consider a {\em class} of approximations to the
effective potential, say $V_{\text{eff}}=V_{\text{triv}}$,
where the shifted fluctuating field
is governed by an effective quadratic hamiltonian.
This includes the one-loop
potential, the gaussian approximation and the infinite set of
post-gaussian calculations where the effective potential
reduces to the sum of a classical background energy and of the zero-point
energy of a massive free field, as at one loop. In this class of
approximations, one finds
\begin{equation}
\label{z2phi}
Z_\varphi= \frac{m^2_H }{V''_{\rm triv}(v_B)} \sim \ln \Lambda
\end{equation}
so that,
to define the physical $v_R$ from the bare $v_B$ one has to apply a non-trivial
correction.  As a consequence,  $v^2_R$
(as now obtained from $v^2_B$ through Eq.~(\ref{vR}) with a
logarithmically divergent $Z_\varphi$ as in Eq.~(\ref{z2phi}))
and $m^2_H\sim g_Rv^2_B$ scale uniformly in the continuum limit.
At the same time, by construction, the shifted field becomes governed by
a quadratic hamiltonian in the continuum limit, so that
one finds $Z_{\text {prop}} \to 1$ as in leading-order perturbation theory.

By adopting this alternative interpretation of `triviality' there are
important phenomenological implications. In fact,
assuming to know the value of $v_R$,
a measurement of $m_H$ would not provide any information on the magnitude
of $\Lambda$ since the ratio
$C=m_H/v_R$ is now a cutoff-independent quantity. Moreover,
in this approach, the quantity $C$ does not represent
the measure of any {\em observable} interaction (see the Conclusions
of Ref.\cite{Agodi:1995qv}).

The difference between $Z_\varphi$ and $Z_{\text{prop}}$
has an important physical meaning, being a
distinctive feature of the
Bose condensation phenomenon~\cite{Consoli:1999ni}.
In the class of `triviality-compatible'
approximations to the effective potential,
one finds
$m_H/v_R=2 \pi \sqrt{2 \zeta}$, with $0< \zeta \leq 2$~\cite{Consoli:1999ni},
$\zeta$
being a cutoff-independent number determined by the quadratic shape of
the effective potential
$V_{\text {eff}}(\varphi_R)$
at $\varphi_R=0$. For instance, $\zeta=1$
corresponds to the classically scale-invariant case or `Coleman-Weinberg
regime'.

As for the standard interpretation of `triviality', the
gaussian effective potential approach can
also be extended to any number N of scalar field components
\cite{Stevenson:1987nb}.
In particular, when studying the continuum limit in the
large-N limit of the theory, one has
to take into account the non-uniformity of the two limits,
cutoff $\Lambda \to \infty$ and $N \to \infty$
\cite{Ritschel:1992ss,Ritschel:1994vr}. This is crucial
to understand the difference with respect to the standard large-N
analysis.

To check the alternative picture of Refs.
\cite{Consoli:1994jr,Consoli:1997ra,Consoli:1999ni}
against the standard point of
view, one can run numerical simulations of the theory and
check the scaling properties of the squared Higgs lattice mass
$m^2_{\text{latt}}$
against those of the inverse
zero-momentum susceptibility $1/\chi_{\text{latt}}$.
 According to perturbation theory, these two quantities should
scale uniformly in the continuum limit. Therefore, a lattice computation of
$Z_\varphi \equiv m^2_{\text{latt}} \chi_{\text{latt}}$
can resolve the issue. If `triviality' is true
and perturbation theory is right, assuming a single rescaling factor
$Z_\varphi\sim Z_{\text {prop}} $,
a lattice simulation has to show unambiguously that
$Z_\varphi$ tends to unity when approaching the continuum limit.
%\begin{equation}
%\label{zlatt}
%Z_\varphi \equiv
%m^2_{\text{latt}} \chi_{\text{latt}}
%\end{equation}

In this respect, we observe that
numerical evidence for different cutoff dependencies of
$Z_\varphi$ and $ Z_{\text{prop}}$ has already been reported in
Refs.~\cite{Cea:1998hy,Cea:1999kn,Cea:1999zu}. In those calculations,
performed in the Ising limit of the one-component theory,
one was fitting the lattice data
for the connected propagator to the (lattice version of the)
two-parameter form
\begin{equation}
\label{gprop}
G_{\text{fit}}(p)= \frac{Z_{\text{prop}}}{ p^2 + m^2_{\text{latt}} }
\end{equation}
After computing the lattice zero-momentum susceptibility
$\chi_{\text{latt}}$,
it was possible to compare the value
$Z_\varphi \equiv m^2_{\text{latt}} \chi_{\text{latt}}$
with the fitted $Z_{\text{prop}}$, both in the symmetric and broken phases.
While no difference was found in the symmetric phase, $Z_\varphi$
and $Z_{\text{prop}}$ were found to be sizeably different
in the broken phase. In particular, $Z_{\text{prop}}$ was very slowly
varying and steadily approaching unity from below in the continuum limit
consistently with K\'allen-Lehmann representation and `triviality'.
$Z_\varphi$, on the other hand,
was found to rapidly increase {\em above} unity in the same limit. The
observed trend was consistent with the logarithmically increasing
trend predicted in
Refs.\cite{Consoli:1994jr,Consoli:1997ra,Consoli:1999ni}.

Now, to our knowledge, with the exception of
Refs.~\cite{Cea:1998hy,Cea:1999kn,Cea:1999zu}, there are no other systematic
investigations of the scaling properties of
$m_{\text{latt}}$ vs. $\chi_{\text{latt}}$ down to values
of lattice mass $m_{\text{latt}} \sim 0.2$. Therefore we might conclude that,
at the present, the alternative theoretical scenario of
Refs.\cite{Consoli:1994jr,Consoli:1997ra,Consoli:1999ni} is
selected by the lattice data.

A possible objection to this conclusion is that the two-parameter form
Eq.~(\ref{gprop}), although providing a good description of
the lattice data, neglects higher-order corrections to the structure
of the propagator. As a consequence, one might object that the
extraction of the various parameters
is affected in an uncontrolled way thus obscuring the observed difference
between $Z_\varphi$ and $Z_{\text{prop}}$.

This objection is not very serious.
In fact, if  `triviality' is true, a two-parameter fit to the
propagator data should become
a better and better approximation approaching the continuum limit
where the genuine perturbative corrections
${\mathcal{O}}(g_R)$ vanish as $\sim 1/\ln\Lambda$.
Therefore, neglected perturbative corrections
that become less and less important
can hardly explain the observed difference between
$Z_\varphi$ and $Z_{\text{prop}}$ that, instead, becomes larger  and larger.

However, to provide additional evidence,
we have decided to change strategy and
perform a new set of lattice calculations of the
zero-momentum susceptibility. In this way,
rather than directly computing the Higgs mass on the lattice,
we shall compare the scaling properties of $\chi_{\rm latt}$ with
the squared mass values predicted by perturbation theory.
Thus, at the same time, we shall be able to check:
i) the previous numerical indications obtained in
Refs.~\cite{Cea:1998hy,Cea:1999kn,Cea:1999zu} and ii)
the internal consistency
of the standard interpretation of `triviality'  that has been
accepted so far. This new computation is consistent with the same
$Z_\varphi\sim \ln \Lambda$ trend observed in
Refs.~\cite{Cea:1998hy,Cea:1999kn,Cea:1999zu} and
will be presented in Sect.~\ref{latticeZphi}. Further evidences are
presented in Sect.~\ref{further} where a different
method to combine the lattice observables leads to the
same conclusion. Finally,
Sect.~\ref{summary} will contain a summary and
a discussion of some general consequences of our results.

\section{The lattice computation of $Z_\varphi$}
\label{latticeZphi}
Our numerical simulations were performed in the Ising limit
that traditionally has been chosen as a convenient laboratory for the
numerical analysis of the theory. In this limit,
a one-component $\Phi^4_4$ theory becomes governed by the lattice
action
\begin{equation}
\label{ising}
S_{\text{Ising}} = -\kappa \sum_x\sum_{\mu} \left[ \phi(x+\hat
e_{\mu})\phi(x) + \phi(x-\hat e_{\mu})\phi(x) \right]
\end{equation}
where $\phi(x)$ takes only the values $\pm 1$.
Using the Swendsen-Wang  and Wolff cluster algorithms we have computed
the bare magnetization:
\begin{equation}
\label{baremagn}
 v_B=\langle |\phi| \rangle \quad , \quad \phi \equiv \sum_x
\phi(x)/L^4
\end{equation}
(where $\phi$  is the average field for each lattice configuration) and
the zero-momentum susceptibility:
\begin{equation}
\label{chi}
 \chi_{\text{latt}}=L^4 \left[ \left\langle |\phi|^2
\right\rangle - \left\langle |\phi| \right\rangle^2 \right] .
\end{equation}
We used different lattice
sizes at each value of $\kappa$ to have a check of the finite-size
effects. Statistical errors have been estimated using the
jackknife. Pseudo-random numbers have been generated using the {\sc Ranlux}
algorithm~\cite{Luscher:1994dy,James:1994vv,Shchur:1998wi} with the highest
possible 'luxury'.
As a check of the goodness of our simulations, we show in Table~1
the comparison with
previous determinations of $\langle |\phi| \rangle$ and
$\chi_{\text {latt}}$
obtained by other authors
~\cite{Jansen:1989cw}.
\begin{table}[t]
\label{table1}
\caption{
We compare our determinations of
$\langle |\phi| \rangle$ and $\chi_{\text{latt}}$ for given
$\kappa$ with corresponding determinations found in the
literature~\cite{Jansen:1989cw}.  In the algorithm column, 'S-W'
stands for the  Swendsen-Wang algorithm, while 'W' stands for the
Wolff algorithm.}
\begin{center}
\begin{tabular}{cccccc}
$\kappa$   &lattice  &algorithm
& $\langle |\phi| \rangle$  &$\chi_{\text{latt}}$    \\
\hline
0.074  &$20^3 \times 24$  &W                          &            &142.21 (1.11)   \\
0.074  &$20^3 \times 24$  &Ref.\cite{Montvay:1987us}  &            &142.6 (8)
\\ \hline
0.077  &$32^4$  &S-W    & 0.38951(1) &18.21(4)   \\
0.077  &$16^4$  &Ref.\cite{Jansen:1989cw}  & 0.38947(2) &18.18(2)
\\ \hline
0.076  &$20^4$  &W   & 0.30165(8) &37.59(31)   \\
0.076  &$20^4$  &Ref.\cite{Jansen:1989cw}    & 0.30158(2) &37.85(6)   \\
\end{tabular}
\end{center}
\end{table}
\begin{table}[t]
\label{table2}
\caption{
The details of the lattice simulations for each $\kappa$ corresponding to $m_{\text{input}}$.
In the algorithm column, 'S-W' stands for the Swendsen-Wang algorithm~\cite{Swendsen:1987ce},
while 'W' stands for the Wolff algorithm~\cite{Wolff:1989uh}. 'Ksweeps' stands for
sweeps multiplied by $10^3$.}
\begin{center}
\begin{tabular}{ccccccc}
$m_{\text{input}}$   &$\kappa$   &lattice  &algorithm &Ksweeps
&$\chi_{\text{latt}}$  &$v_B=\langle |\phi| \rangle$\\
\hline
0.4     &0.0759     &$32^4$  &S-W  & 1750  &41.714 (0.132)    & 0.290301 (21)                \\
0.4     &0.0759     &$48^4$  &W    &   60  &41.948 (0.927)    & 0.290283 (52)   \\ \hline
0.35    &0.075628   &$48^4$  &W    &  130  &58.699 (0.420)    & 0.255800 (18)   \\ \hline
0.3     &0.0754     &$32^4$  &S-W  &  345  &87.449 (0.758)    & 0.220540 (75)                \\
0.3     &0.0754     &$48^4$  &W    &  406  &87.821 (0.555)    & 0.220482 (19)   \\ \hline
0.275   &0.075313   &$48^4$  &W    &   53  &104.156 (1.305)   & 0.204771 (40)   \\ \hline
0.25    &0.075231   &$60^4$  &W    &   42  &130.798 (1.369)   & 0.188119 (31)   \\ \hline
0.2     &0.0751     &$48^4$  &W    &   27  &203.828 (3.058)   & 0.156649 (103)                \\
0.2     &0.0751     &$52^4$  &W    &   48  &201.191 (6.140)   & 0.156535 (65)                \\
0.2     &0.0751     &$60^4$  &W    &    7  &202.398 (8.614)   & 0.156476 (15)   \\ \hline
0.15    &0.074968   &$68^4$  &W    &   25  &460.199 (4.884)   & 0.112611 (51)   \\ \hline
0.1     &0.0749     &$68^4$  &W    &   24  &1125.444 (36.365) & 0.077358 (123)                \\
0.1     &0.0749     &$72^4$  &W    &    8  &1140.880 (39.025) & 0.077515 (210)  \\
\end{tabular}
\end{center}
\end{table}
As anticipated, rather than computing the Higgs mass on the lattice
as in Refs.~\cite{Cea:1998hy,Cea:1999kn,Cea:1999zu},
we shall use the perturbative predictions for its value and adopt the
L\"uscher-Weisz scheme~\cite{Luscher:1988ek}.
To this end, we shall denote by
$m_{\text{input}}$ the value of the parameter $m_R$ reported in
the first column of Table~3 in Ref.~\cite{Luscher:1988ek} for any
value of $\kappa$ (the { Ising limit corresponding to the value of
the other parameter $\bar{\lambda}=1$).

Our data for $\chi_{\text {latt}}$
at various $\kappa$ are reported in Table~2
for the range
$0.1 \leq m_{\text{input}}\leq 0.4$
(the relevant $\kappa$'s for $m_{\text{input}}=0.15, 0.25, 0.275, 0.35$ have been
determined through a numerical interpolation of the data shown in the
L\"uscher-Weisz Table).
\begin{figure}[t]
\begin{center}
%\framebox{
\includegraphics[width=0.8\textwidth,clip]{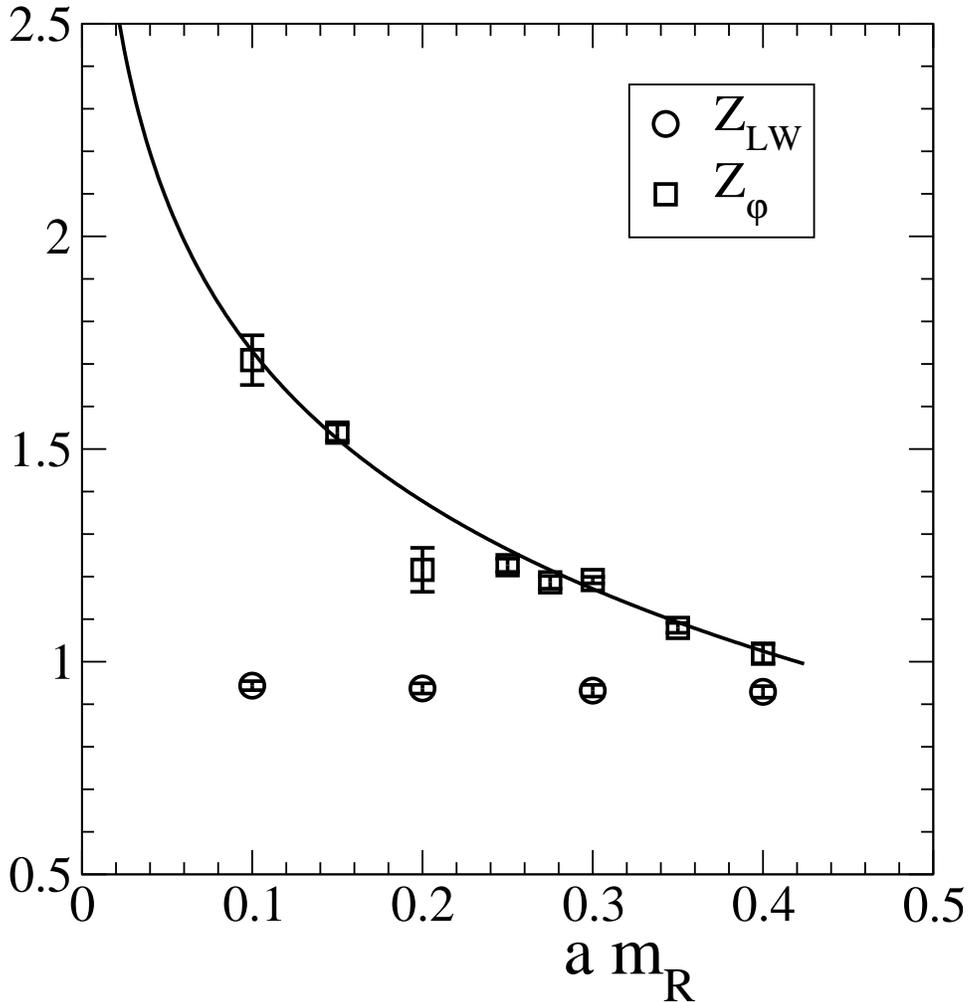}
\caption{The lattice data for $Z_\varphi$, as defined in
Eq.~(\ref{zphi}), and its perturbative prediction $Z_{\text{LW}}$
versus $m_{\text{input}}=a m_R$. The solid line is a fit to the form
Eq.~(\ref{fit-form}) with $B=0.50$.}
\end{center}
\end{figure}
\begin{table}[t]
\label{table3}
\caption{
The product $v^2_B \chi_{\text{latt}}$ versus $\kappa$.
These values correspond, for each $\kappa$, to the most precise entries reported
in Table~2.}
\begin{center}
\begin{tabular}{cc}
$\kappa$   &$v^2_B \chi_{\text{latt}}$ \\
\hline
   0.075900   & 3.5154   ~(122) \\
   0.075628   & 3.8409   ~(275) \\
   0.075400   & 4.2692   ~(270) \\
   0.075313   & 4.3674   ~(547) \\
   0.075231   & 4.6288   ~(485) \\
   0.075100   & 4.9907   ~(755) \\
   0.074968   & 5.8359   ~(622) \\
   0.074900   & 6.7349   (2186) \\
\end{tabular}
\end{center}
\end{table}
\begin{figure}[t]
\begin{center}
%\framebox{
\includegraphics[width=0.9\textwidth,clip]{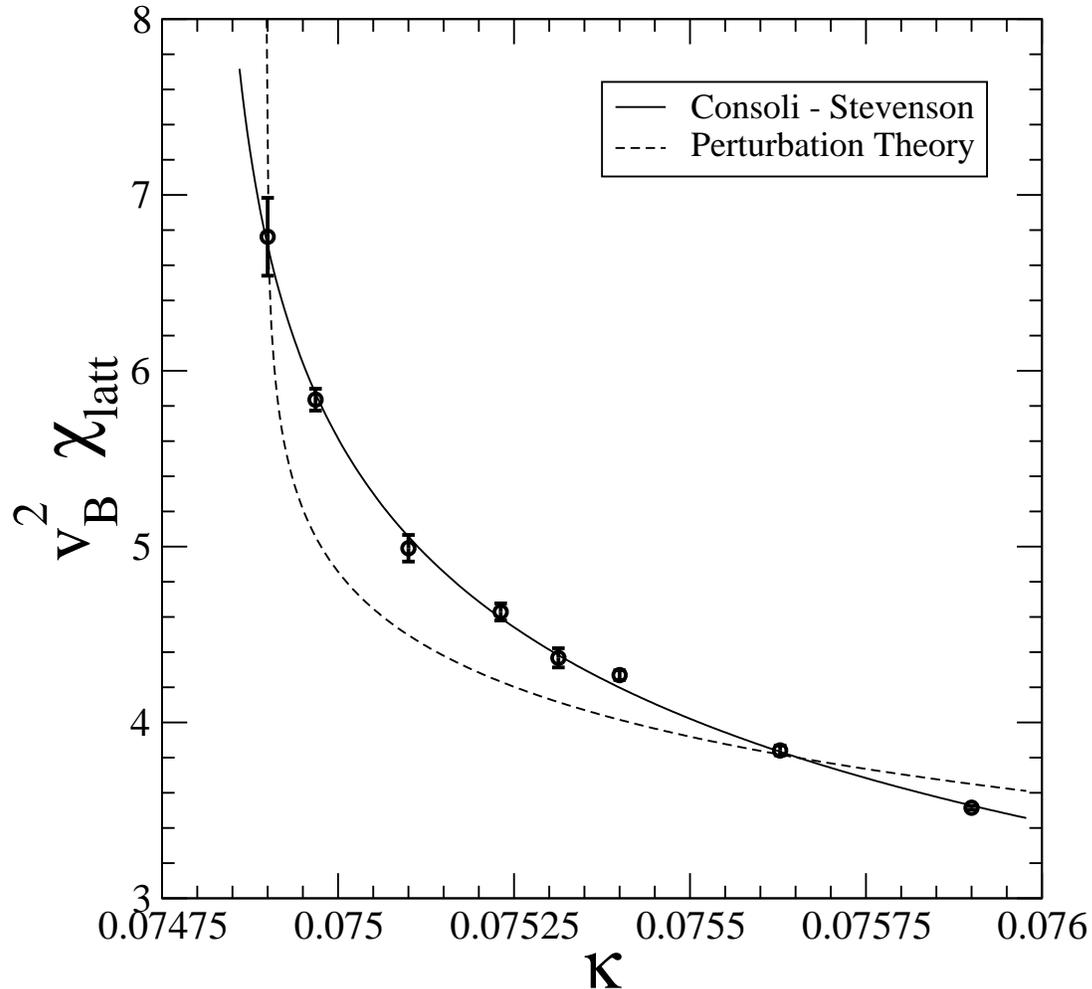}
\caption{We show the lattice data for $v^2_B\chi_{\text{latt}}$
of Table 3 together with the corresponding fit Eq.~(\ref{Ffit}) for
$\gamma=2$  (Consoli-Stevenson) and $\gamma=1$ (Perturbation Theory).}
\end{center}
\end{figure}
At this point, we can compare the quantity
\begin{equation}
\label{zphi}
       Z_\varphi\equiv 2\kappa m^2_{\text{input}} \chi_{\text{latt}}
\end{equation}
with the perturbative determination
\begin{equation}
\label{zlw}
Z_{\text{LW}}\equiv 2\kappa Z_R
\end{equation}
where $Z_R$ is defined in the third column of Table~3 in
Ref.~\cite{Luscher:1988ek}.

The values of $Z_\varphi$ and
$Z_{\text{LW}}$ are reported in Fig.~1.
As one can check, the two $Z$'s follow completely different trends
and the discrepancy becomes larger and
larger when approaching the continuum limit, precisely the same
behavior observed in
Refs.\cite{Cea:1998hy,Cea:1999kn,Cea:1999zu}.
We also fitted the values for $Z_\varphi$ to the form ($\Lambda=\pi/a$)
\begin{equation}
\label{fit-form}
Z_\varphi= B \,  \ln{(\Lambda/m_R)} \,.
\end{equation}
Notice that the lattice data
are completely consistent with the prediction
$Z_\varphi \sim \ln \Lambda$ of
Refs.\cite{Consoli:1994jr,Agodi:1995qv,Consoli:1997ra,Consoli:1999ni}.

Of course, one might object that the discrepancy between
lattice data and the perturbative $Z_{\rm LW}$ depends on
restricting, for each given
$m_{\rm input}$, to the central values of $\kappa$ in the
L\"uscher-Weisz table. As a matter of
fact, there is a theoretical uncertainty in $\kappa$, for each given
$m_{\rm input}$, that might be taken into account when comparing with the
perturbative $Z_{\rm LW}$'s. For instance, for
$m_{\rm input}=0.3$, where the $\kappa$-range predicted
in Ref.~\cite{Luscher:1988ek} is $0.0754(2)$, one might also compute
$\chi_{\rm latt}$ for
$\kappa=0.0756$ (obtaining a value
$\chi_{\rm latt}\sim 60$) and for
$\kappa=0.0752$ (obtaining a value
$\chi_{\rm latt}\sim 130$).
As a consequence, for $m_{\rm input}=0.3$, one might
also conclude that $Z_\varphi$, as defined
in Eq.(\ref{zphi}), lies in the range $0.8-1.7$, consistently with
the prediction of
Ref.~\cite{Luscher:1988ek}
$Z_{\rm LW}=0.932(13)$.

Adopting this point of view does not represent, however,
a solution of the problem. In fact, by inspection of Fig.~1, one can see that
the range of $\kappa$-values for which $Z_\varphi$,
as defined in Eq.(\ref{zphi}), is still
consistent with the perturbative $Z_{\rm LW}$ becomes smaller and smaller
when approaching the continuum limit. As a matter of fact, this range
vanishes for $m_{\rm input} \leq 0.08$.
This can easily be checked noticing that
for $m_{\rm input}=0.08$
Ref.~\cite{Luscher:1988ek} predicts $\kappa=0.07481(8)$, i.e.
{\em smaller} than $0.0749$. Therefore, by inspection
of our Table 2, the lattice susceptibility will be
definitely {\em larger} than its value for $\kappa=0.0749$,
$\chi_{\rm latt}\sim 1100$, so that
using Eq.(\ref{zphi}), one gets the {\em lower bound}
$Z_\varphi > 1.05$. This cannot be reconciled with the perturbative prediction,
for $m_{\rm input}=0.08$, $Z_{\rm LW}=0.947(10)$.

In our opinion, one should not
ignore this discrepancy, for instance by attempting to further enlarge
the error bars of the perturbative predictions so as to reach a
marginal consistency with the lattice data. In fact, in this way no
meaningful test of perturbation theory will ever be possible. This is in
contrast with the situation for the symmetric phase where
lattice data and theoretical predictions based on the
central values of $\kappa$, as given
in Table~3 of Ref.~\cite{Luscher:1987ay}, agree to good accuracy,
see Refs.~\cite{Cea:1998hy,Cea:1999kn,Cea:1999zu}. As an additional check,
we have computed the
zero-momentum susceptibility for the central value
$\kappa=0.0741$ that corresponds to
$m_{\text{input}}=0.2$. From our
result on a $32^4$ lattice $\chi_{\text{latt}}=161.94 \pm 0.67$,
using again Eq.~(\ref{zphi}), we obtain $Z_\varphi=0.960\pm 0.004$ in
good agreement with the corresponding L\"uscher-Weisz prediction
$Z_{\text{LW}}=0.975\pm 0.010$. This is also consistent with the picture of
Refs.\cite{Consoli:1994jr,Consoli:1997ra,Consoli:1999ni}
where the deviations from perturbation theory are only due to the presence
of the scalar condensate.
\section{Further lattice determination of $Z_\varphi$}
\label{further}

There is, however, a method to analyze the lattice data where
the theoretical uncertainty in the $\kappa-m_R$ correlation can be
eliminated. In fact, one can perform a fit of the lattice data with the
functional form expected in the standard perturbative approach or with the
functional form expected in the alternative scenario of
Refs.\cite{Consoli:1994jr,Consoli:1997ra,Consoli:1999ni}. If this is done,
the lattice data confirm a logarithmically divergent $Z_\varphi$ in agreement
with the previous indications from Fig.~1.

To this end, let us observe that, both in perturbation theory and according to
Refs.~\cite{Consoli:1994jr,Consoli:1997ra,Consoli:1999ni}, the bare
squared field expectation value $v^2_B$ is predicted to
diverge in units of the physical Higgs boson mass as
\begin{equation}
\label{v2Blog}
\frac{v^2_B}{m^2_R} \sim  \ln{(\Lambda/m_R)} \,.
\end{equation}
Therefore, in the perturbative approach, where
the zero-momentum susceptibility
is predicted to scale uniformly with the inverse squared Higgs mass, i.e.
$\chi_2(0)m^2_R\sim 1$, one expects (PT=Perturbation Theory)
\begin{equation}
\label{v2Bchi2pt}
\left[ v^2_B \chi_2(0) \right]_{\text{PT}}
\sim  \ln{(\Lambda/m_R)} \,.
\end{equation}
On the other hand, in the approach of
Refs.~\cite{Consoli:1994jr,Consoli:1997ra,Consoli:1999ni} one
predicts
$Z_\varphi=\chi_2(0)m^2_R \sim \ln{(\Lambda/m_R)}$ so that, in this case,
one would rather expect (CS=Consoli-Stevenson)
\begin{equation}
\label{v2Bchi2cs}
\left[ v^2_B \chi_2(0) \right]_{\text{CS}}
\sim  \ln^2{(\Lambda/m_R)}   \,.
\end{equation}
The two predictions in
Eq.(\ref{v2Bchi2pt}) and in Eq.(\ref{v2Bchi2cs}) are free of
theoretical uncertainties due to the $\kappa-m_R$ correlation and can be directly
compared with the lattice data for the product $v^2_B \chi_2(0)$ reported
in our Table~3. These data can be fitted to the 3-parameter form
\begin{equation}
\label{Ffit}
\alpha |\ln(\kappa-\kappa_c)|^\gamma
\end{equation}
where $\alpha$ is a normalization constant and
we shall set the exponent $\gamma=1$, according to
Eq.(\ref{v2Bchi2pt}), or $\gamma=2$ according
to Eq.(\ref{v2Bchi2cs}).

Now, fixing
$\gamma=1$ one obtains the totally unacceptable value chi-square$\simeq496$
 for 6 degrees of freedom (see Fig.~2). On the other hand,
fixing $\gamma=2$, one obtains a good fit of the lattice data
(chi-square$\simeq9$ for 6 degrees of freedom) with a rather precise
determination of the critical point from the broken-symmetry phase
$\kappa_c=0.074818(11)$ consistently with the estimate
$\kappa_c=0.074834(15)$ from the symmetric phase~\cite{Gaunt:1979aa}.
This conclusion is also confirmed by the results of the full
3-parameter fit, $\gamma=2.20^{+0.24}_{-0.21}$ and
$\kappa_c=0.07478(6)$ with a chi-square of 6.4 for 5 degrees of freedom.
Therefore, between the two alternatives Eq.(\ref{v2Bchi2pt}) and
Eq.(\ref{v2Bchi2cs}),
the lattice data prefer unambiguously the theoretical scenario of
Refs.~\cite{Consoli:1994jr,Consoli:1997ra,Consoli:1999ni}.

Before concluding, a brief comment about the L\"uscher-Weisz prediction for
the product $v^2_B \chi_2(0)$. This is based on the relation
\begin{equation}
\label{v2bchi2lw}
\left[ v^2_B \chi_2(0) \right]_{\text{LW}} = \frac{3 Z_R^2}{g_R}
\end{equation}
where $Z_R$ and $g_R$ are given in Table 3 of
Ref.~\cite{Luscher:1988ek} as functions of the mass parameter $m_R$.

Therefore, a comparison of Eq.(\ref{v2bchi2lw}) with the lattice data
reported in our Table 3 re-introduces unavoidably the
uncertainties associated with the $\kappa-m_R$ correlation. For instance, for
$\kappa=0.0759$ the prediction of Ref.~\cite{Luscher:1988ek} would be
$\left[ v^2_B \chi_2(0) \right]_{\text{LW}}=4.16$ to be compared with the
value $3.5154(122)$ in our Table 3. On the other hand, if one takes into
account that the relevant values of $Z_R$ and $g_R$ in Eq.(\ref{v2bchi2lw}) are
actually defined
for the mass value $m_R=0.4$, for which the entire range is $\kappa=0.0759(2)$, one
might also conclude that the theoretical prediction lies in the range
$3.76\leq \left[ v^2_B \chi_2(0) \right]_{\text{LW}}\leq 4.63$ that, with a
short-hand notation, can be expressed as
$\left[ v^2_B \chi_2(0) \right]_{\text{LW}}= 4.20\pm 0.43$.

Notice, however, that the difference between the prediction
$4.20\pm 0.43$ and the lattice
value $3.5154(122)$ should not be considered a modest $1.6\sigma$
discrepancy. In fact the interval $3.76-4.63$ represents the
{\it entire theoretical range}
allowed by the numerical solution of the renormalization-group
equations. At the same time, we observe that
the agreement does not seem to improve approaching
the continuum limit. For instance, for $m_R=0.2$, where the range is
$\kappa=0.0751(1)$, Ref.~\cite{Luscher:1988ek} predicts
$5.42\leq \left[ v^2_B \chi_2(0) \right]_{\text{LW}}\leq 6.31$
to be compared with the lattice
value $4.9907(755)$ in our Table 3. This situation
might be reminiscent of the discrepancy in the value of $Z_R$,
pointed out by Jansen et al. ~\cite{Jansen:1989cw}, that was not improving
when approaching the continuum limit.

\section{Summary and outlook}
\label{summary}
In this paper we have reported the results of a lattice simulation
dedicated to check the first numerical indications
~\cite{Cea:1998hy,Cea:1999kn,Cea:1999zu} for a
logarithmically divergent rescaling of the scalar condensate.
Two different methods, see Fig.~1 and Fig.~2, support
the same behavior $Z_\varphi\sim \ln \Lambda$ predicted in
Refs.~\cite{Consoli:1994jr,Consoli:1997ra,Consoli:1999ni} from the analysis
of the effective potential.

As anticipated in the Introduction, within this scenario
there is one substantial phenomenological implication
that concerns the relative scaling of
the physical Higgs mass $m_H$ and of
the physical $v_R$ for which
$V''_{\text{eff}}(v_R)=m^2_H$. Namely, once $v^2_R$
is computed from the bare $v^2_B$
through Eq.(\ref{vR}) with a value $Z=Z_\varphi\sim B\ln \Lambda$, so that
\begin{equation}
v^2_R\sim v^2_B~\frac{1}{B\ln \Lambda}
\end{equation}
{\em this} $v^2_R$ will scale uniformly
with
\begin{equation}
m^2_H \sim v^2_B~ \frac{A}{\ln \Lambda}
\end{equation}
(where we have set $g_R\sim \frac{A}{\ln\Lambda}$).

Therefore, if $v_R$ is identified with a physical scale (e.g. $v_R\sim 246$ GeV),
there are {\em no} upper bounds on $m_H$ from `triviality' since
\begin{equation}
C=\frac{m_H}{v_R}\sim \sqrt{AB}
\end{equation}
is a {\em cutoff-independent} quantity~\cite{Cea:2003gp}.
The numerical value of $C$, however,
could depend on the direction chosen to approach the critical
line in the more general two-parameter $(\kappa,\bar{\lambda})$ form
~\cite{Luscher:1988ek} of $\Phi^4$ theory.
Therefore, a new set of lattice simulations is needed to compute
the zero-momentum susceptibility $\chi_{\rm latt}$ outside of the Ising limit
$\bar{\lambda}=1$.

\vskip 30 pt
\centerline{\bf{ACKNOWLEDGEMENTS}}
We thank P. M. Stevenson for many useful discussions and collaboration.

\providecommand{\href}[2]{#2}\begingroup\raggedright\endgroup


\begin{thebibliography}{10}

\bibitem{Sokal_book}
R.~Fernandez, J.~Fr{\"o}hlich, and A.~D. Sokal, {\em {Random Walks, Critical
  Phenomena, and Triviality in Quantum Field Theory}}.
\newblock Springer-Verlag, Berlin, 1992.

\bibitem{Lang:1993sy}
C.~B. Lang, {\it Computer stochastics in scalar quantum field theory, in
  {proceedings of NATO Advanced Study Institute on Stochastic Analysis and
  Applications in Physics, Funchal, Madeira, 6-19 Aug 1993}},
  \href{http://xxx.lanl.gov/abs/hep-lat/9312004}{{\tt hep-lat/9312004}}.

\bibitem{Kenna:2004cm}
R.~Kenna, {\it Finite size scaling for {O(N) $\phi^4$} theory at the upper
  critical dimension},  \href{http://xxx.lanl.gov/abs/hep-lat/0405023}{{\tt
  hep-lat/0405023}}.

\bibitem{Consoli:1994jr}
M.~Consoli and P.~M. Stevenson, {\it The non trivial effective potential of the
  'trivial' {$\lambda \phi^4$} theory: a lattice test},  {\em Z. Phys.} {\bf
  C63} (1994) 427--436,
  [\href{http://xxx.lanl.gov/abs/http://arXiv.org/abs/hep-ph/9310338}{{\tt
  http://arXiv.org/abs/hep-ph/9310338}}].

\bibitem{Consoli:1997ra}
M.~Consoli and P.~M. Stevenson, {\it Model-dependent field renormalization and
  triviality in {$\lambda \phi^4$} theory},  {\em Phys. Lett.} {\bf B391}
  (1997) 144--149.

\bibitem{Consoli:1999ni}
M.~Consoli and P.~M. Stevenson, {\it Physical mechanisms generating spontaneous
  symmetry breaking and a hierarchy of scales},  {\em Int. J. Mod. Phys.} {\bf
  A15} (2000) 133,
  [\href{http://xxx.lanl.gov/abs/http://arXiv.org/abs/hep-ph/9905427}{{\tt
  http://arXiv.org/abs/hep-ph/9905427}}].

\bibitem{Agodi:1995qv}
A.~Agodi, G.~Andronico, and M.~Consoli, {\it Lattice {$\Phi^4_4$} effective
  potential giving spontaneous symmetry breaking and the role of the {Higgs}
  mass},  {\em Z. Phys.} {\bf C66} (1995) 439--452,
  [\href{http://xxx.lanl.gov/abs/hep-lat/9410001}{{\tt hep-lat/9410001}}].

\bibitem{Stevenson:1987nb}
P.~M. Stevenson, B.~Alles, and R.~Tarrach, {\it {O(N)} symmetric {$\lambda
  \varphi^4$} theory: the {Gaussian} effective potential approach},  {\em Phys.
  Rev.} {\bf D35} (1987) 2407.

\bibitem{Ritschel:1992ss}
U.~Ritschel, I.~Stancu, and P.~M. Stevenson, {\it Unconventional large {N}
  limit of the {Gaussian} effective potential and the phase transition in
  {$\lambda \varphi^4$} theory},  {\em Z. Phys.} {\bf C54} (1992) 627--634.

\bibitem{Ritschel:1994vr}
U.~Ritschel, {\it {NonGaussian} corrections to {Higgs} mass in autonomous
  {$\lambda \varphi^4$} in (3+1)-dimensions},  {\em Z. Phys.} {\bf C63} (1994)
  345--350, [\href{http://xxx.lanl.gov/abs/hep-ph/9210206}{{\tt
  hep-ph/9210206}}].

\bibitem{Cea:1998hy}
P.~Cea, M.~Consoli, and L.~Cosmai, {\it First lattice evidence for a
  non-trivial renormalization of the {Higgs} condensate},  {\em Mod. Phys.
  Lett.} {\bf A13} (1998) 2361--2368,
  [\href{http://xxx.lanl.gov/abs/http://arXiv.org/abs/hep-lat/9805005}{{\tt
  http://arXiv.org/abs/hep-lat/9805005}}].

\bibitem{Cea:1999kn}
P.~Cea, M.~Consoli, L.~Cosmai, and P.~M. Stevenson, {\it Further lattice
  evidence for a large re-scaling of the {Higgs} condensate},  {\em Mod. Phys.
  Lett.} {\bf A14} (1999) 1673--1688,
  [\href{http://xxx.lanl.gov/abs/http://arXiv.org/abs/hep-lat/9902020}{{\tt
  http://arXiv.org/abs/hep-lat/9902020}}].

\bibitem{Cea:1999zu}
P.~Cea, M.~Consoli, and L.~Cosmai, {\it Large rescaling of the {Higgs}
  condensate: Theoretical motivations and lattice results},  {\em Nucl. Phys.
  Proc. Suppl.} {\bf 83} (2000) 658--660,
  [\href{http://xxx.lanl.gov/abs/hep-lat/9909055}{{\tt hep-lat/9909055}}].

\bibitem{Luscher:1994dy}
M.~{L\"uscher}, {\it A portable high quality random number generator for
  lattice field theory simulations},  {\em Comput. Phys. Commun.} {\bf 79}
  (1994) 100--110, [\href{http://xxx.lanl.gov/abs/hep-lat/9309020}{{\tt
  hep-lat/9309020}}].

\bibitem{James:1994vv}
F.~James, {\it Ranlux: A fortran implementation of the high quality
  pseudorandom number generator of {L\"uscher}},  {\em Comp. Phys. Commun.}
  {\bf 79} (1994) 111--114.

\bibitem{Shchur:1998wi}
L.~N. Shchur and P.~Butera, {\it The {\sc ranlux} generator: Resonances in a
  random walk test},  {\em Int. J. Mod. Phys.} {\bf C9} (1998) 607--624,
  [\href{http://xxx.lanl.gov/abs/hep-lat/9805017}{{\tt hep-lat/9805017}}].

\bibitem{Jansen:1989cw}
K.~Jansen, T.~Trappenberg, I.~Montvay, G.~Munster, and U.~Wolff, {\it Broken
  phase of the four-dimensional {Ising} model in a finite volume},  {\em Nucl.
  Phys.} {\bf B322} (1989) 698.

\bibitem{Montvay:1987us}
I.~Montvay and P.~Weisz, {\it Numerical study of finite volume effects in the
  four- dimensional ising model},  {\em Nucl. Phys.} {\bf B290} (1987) 327.

\bibitem{Swendsen:1987ce}
R.~H. Swendsen and J.-S. Wang, {\it Nonuniversal critical dynamics in {Monte
  Carlo} simulations},  {\em Phys. Rev. Lett.} {\bf 58} (1987) 86--88.

\bibitem{Wolff:1989uh}
U.~Wolff, {\it Collective {Monte Carlo} updating for spin systems},  {\em Phys.
  Rev. Lett.} {\bf 62} (1989) 361.

\bibitem{Luscher:1988ek}
M.~{L\"uscher} and P.~Weisz, {\it Scaling laws and triviality bounds in the
  lattice {$\phi^4$} theory. 2. {One} component model in the phase with
  spontaneous symmetry breaking},  {\em Nucl. Phys.} {\bf B295} (1988) 65.

\bibitem{Luscher:1987ay}
M.~{L\"uscher} and P.~Weisz, {\it Scaling laws and trivality bounds in the
  lattice {$\phi^4$} theory. 1. {One} component model in the symmetric phase},
  {\em Nucl. Phys.} {\bf B290} (1987) 25.

\bibitem{Gaunt:1979aa}
D.~S. Gaunt, M.~F. Sykes, and S.~McKenzie {\em J. Phys.} {\bf A12} (1979) 871.

\bibitem{Cea:2003gp}
P.~Cea, M.~Consoli, and L.~Cosmai, {\it Indications on the {Higgs} boson mass
  from lattice simulations},  {\em Nucl. Phys. Proc. Suppl.} {\bf 129} (2004)
  780--782, [\href{http://xxx.lanl.gov/abs/hep-lat/0309050}{{\tt
  hep-lat/0309050}}].

\end{thebibliography}
\end{document}